\documentclass[showpacs,prl,aps,10pt,superscriptaddress,altaffilletter,floatfix,twocolumn]{revtex4-1}
\usepackage{graphicx}
\usepackage{hyperref}
\usepackage{amsmath}
\usepackage{amssymb}
\usepackage{color}
\usepackage{xcolor}
\usepackage{xspace}
\usepackage{dcolumn}
\usepackage{soul}
\usepackage{comment}
\newcolumntype{C}[1]{>{\centering\let\newline\\\arraybackslash\hspace{0pt}}m{#1}}
\usepackage[section]{placeins}
\usepackage[normalem]{ulem} 

\RequirePackage{lineno}

\usepackage{upgreek}

\begin{document}

\newcommand{\isot}[2]{$^{#2}$#1}
\newcommand{\isotbold}[2]{$^{\boldsymbol{#2}}$#1}
\newcommand{\xeiso}{\isot{Xe}{136}\xspace}
\newcommand{\thsrc}{\isot{Th}{228}\xspace}
\newcommand{\cosrc}{\isot{Co}{60}\xspace}
\newcommand{\rasrc}{\isot{Ra}{226}\xspace}
\newcommand{\cssrc}{\isot{Cs}{137}\xspace}
\newcommand{\betascale}  {$\beta$-scale}
\newcommand{\kevkgyr}  {keV$^{-1}$ kg$^{-1}$ yr$^{-1}$}
\newcommand{\nonubb}  {$0\nu \beta\!\beta$\xspace}
\newcommand{\nonubbbf}  {$\boldsymbol{0\nu \beta\!\beta}$\xspace}
\newcommand{\twonubb} {$2\nu \beta\!\beta$\xspace}
\newcommand{\bb} {$\beta\!\beta$\xspace}
\newcommand{\vadc} {ADC$_\text{V}$}
\newcommand{\uadc} {ADC$_\text{U}$}
\newcommand{\mus} {\textmu{}s}
\newcommand{\chisq} {$\chi^2$}
\newcommand{\mum} {\textmu{}m}
\newcommand{\red}[1]{{\xspace\color{red}#1}}
\newcommand{\blue}[1]{{\xspace\color{blue}#1}}
\newcommand{\RunTwoA}{Run 2a}
\newcommand{\RunTwo}{Run 2}
\newcommand{\RunTwoBC}{Runs 2b and 2c}
\newcommand{\SP}[1]{\textsuperscript{#1}}
\newcommand{\SB}[1]{\textsubscript{#1}}
\newcommand{\SPSB}[2]{\rlap{\textsuperscript{#1}}\SB{#2}}
\newcommand{\pmasy}[3]{#1\SPSB{$+$#2}{$-$#3}}
\newcommand{\matel}{$M^{2\nu}$}
\newcommand{\psfac}{$G^{2\nu}$}
\newcommand{\tbeta}{T$_{1/2}^{0\nu\beta\beta}$}
\newcommand{\exolimit}[1][true]{\pmasy{2.6}{1.8}{2.1}$ \cdot 10^{25}$}
\newcommand{\exomeasurement}{\tbeta{}= \exolimit{}~yr}
\newcommand{\U}{\text{U}}
\newcommand{\V}{\text{V}}
\newcommand{\X}{\text{X}}
\newcommand{\Y}{\text{Y}}
\newcommand{\Z}{\text{Z}}
\newcommand{\bqcm}{${\rm Bq~m}^{-3}$}
\newcommand{\nonunorm}{N_{{\rm Err, } 0\nu\beta\beta}}
\newcommand{\nonunum}{n_{0\nu\beta\beta}}
\newcommand{\cussim}[1]{$\sim$#1}
\newcommand{\halflife}[1]{$#1\cdot10^{25}$~yr}
\newcommand{\numspec}[3]{$N_{^{#2}\mathrm{#1}}=#3$}
\newcommand{\TD}[1]{\textcolor{red}{#1}}
\newcommand{\PI}{Phase~I\xspace}
\newcommand{\PII}{Phase~II\xspace}
\newcommand{\Rn}{radon\xspace}
\newcommand\Tstrut{\rule{0pt}{2.6ex}} 

\newcommand\newtext[1]{{\color{red}#1}}
\setstcolor{blue}

\title{The next frontiers for magnetic monopole searches}

\author{O.~Gould}\affiliation{University of Nottingham, Nottingham, UK}
\author{I.~Ostrovskiy}\email[Corresponding author: ]{iostrovskiy@ua.edu}\affiliation{Department of Physics and Astronomy, University of Alabama, Tuscaloosa, Alabama, USA}
\author{A.~Upreti}\affiliation{Department of Physics and Astronomy, University of Alabama, Tuscaloosa, Alabama, USA}

\date{\today}

\begin{abstract}
Magnetic monopoles (MMs) are well-motivated hypothetical particles whose discovery would symmetrize Maxwell equations, explain quantization of electric charge, and probe the gauge structure of the unified theory. Recent models predict MMs with low masses, reinvigorating searches at colliders. However, most theories predict composite MMs, whose production in parton-parton collisions is expected to be suppressed. The Schwinger process, whereby MM pairs tunnel through the vacuum barrier in the presence of a strong magnetic field, is not subject to this limitation. Additionally, the Schwinger cross section can be calculated nonperturbatively. Together, these make it a golden channel for low-mass MM searches. We investigate the Schwinger production of MMs in heavy-ion collisions at future colliders, in collisions of cosmic rays with the atmosphere, and in decay of magnetic fields of cosmic origin. We find that a next-generation collider would provide the best sensitivity, allowing one to discover or exclude MMs with TeV-scale masses. At the same time, exploiting the infrastructure of industrial ore extraction and Antarctic ice drilling could advance the field at a faster timescale and with only a modest investment. In particular, we show with detailed calculations that the proposed experiments will be sensitive to fluxes of low-mass MMs as low as a few units of 10$^{-22}$ cm$^{-2}$s$^{-1}$sr$^{-1}$ in a wide range of Lorentz factors. We also propose deploying dedicated MM detectors in conjunction with cosmic ray observatories to directly investigate if the unexplained, highest energy cosmic rays are MMs. Together, the proposed efforts would define the field of MM searches in the next decades.
\end{abstract}


\maketitle
\section{Introduction}
The magnetic monopole (MM) is a hypothetical particle that carries isolated magnetic charge. It was postulated to exist by Dirac to explain the apparent quantization of the electric charge~\cite{Dirac:1931}. Dirac calculated the fundamental magnetic charge, called Dirac charge, to be:
\begin{equation}
g_{\mathrm{D}} = \frac{e}{2\alpha} \simeq 68.5e,
\end{equation}
where $e$ is the proton charge and $\alpha$ is the fine-structure constant. 

Dirac MMs are elementary particles with no internal structure. In contrast, solutions with isolated magnetic charge that appear in all variants of grand unified theories (GUT) that incorporate electromagnetism~\cite{Hooft:74, Polyakov:1974ek} are composite objects -- a bound state of carriers of the unified and electroweak interactions and other particles~\cite{patrici:2015, vaso_ask_1}. 
The mass of a Dirac MM is a free parameter, while GUT MMs have masses on the order of the GUT scale, i.e., $\sim$10$^{13}$ TeV/c$^2$. However, in models with several stages of symmetry breaking the MM mass is decreased accordingly. Such MMs could be produced after the inflationary epoch and would not catalyze proton decay~\cite{Kephart:2001,pdg:2014}, evading the astrophysical limits on this process. Notably, composite finite-energy MM solutions have recently been discovered in several beyond-the-standard-model field theories, with masses as low as $\sim10^{0}$ TeV/c$^2$~\cite{CHO:1997,cho:2015,Ellis:2016,mavromatos:2017,Arunasalam:2017,Mavromatos:2018,Hung:2020, bps_cho_mason_2018}. Unlike singly charged Dirac MMs, the fundamental magnetic charge predicted by theories based on spontaneously broken gauge symmetries may be $n$ times larger than the Dirac charge, where $n$ is an integer that depends on the global structure of the underlying symmetry group~\cite{Corrigan:1976wk, Tong:2017oea}. For example, in the trinification model, where the MM only carries the U(1) and not color magnetic charges, the fundamental magnetic charge is three units of Dirac charge~\cite{Shafi:2021}. In another instance, the Cho-Mason MM carries two units of Dirac charge~\cite{CHO:1997}. String theories also contain MMs with masses that depend on the string scale, potentially much smaller than the GUT scale~\cite{Wen-Witten:1985}.  

The proliferation of models suggesting low MM masses spurred experimental searches at the LHC. In the last few years, searches for production of MMs in p-p collisions were performed there by ATLAS~\cite{atlas_2012,atlas_2016,atlas_2020,ATLAS2023_MM} and MoEDAL~\cite{moedal_first,moedal_prl:2017,moedal_prl2019,moedal_run2_2, HECO_1, HECO_2}. Predicting the rate and kinematics of MM production is difficult because MMs couple strongly to photons, and so perturbative quantum field theory does not apply, unless appropriate resummation schemes are used~\cite{PhysRevD.100.096005, vaso_ask_2}. Consequently, the leading-order cross section calculations for the assumed Drell-Yan or photon-fusion mechanisms, which are used by the searches, can only be treated as indicative, and the corresponding mass limits are only suited for relative comparisons between experiments. Additionally, the quoted searches concentrate on the production of point-like MMs. This is because the production of composite MMs is expected to be suppressed by a huge factor of e$^{-4/\mathrm{\alpha}}$ in collisions of elementary particles~\cite{Witten:1979,drukier} due to negligible overlap between the initial and final states. A recently proposed approach to search for MM production in collisions of cosmic rays (CRs) with the atmosphere~\cite{ryan_2022} is subject to the same limitation. 

A different production method that overcomes the above limitations is the electromagnetic dual of the Schwinger mechanism~\cite{schwinger,Heisenberg, Sauter}, which describes electron-positron pairs tunnelling through the vacuum barrier in the presence of strong electric fields. As was shown~\cite{affleck,arttu_prd2021_103}, MMs could similarly be produced by short-lived strong magnetic fields created when relativistic heavy ions pass by each other. Importantly, the finite size of composite MMs only enhances the production rate~\cite{arttu_prd2021_103,arttu_prd2020}, and the rate can be calculated nonperturbatively~\cite{arttu_prd2019,arttu_prd2020}. Additionally, it was shown recently that MMs could have been produced via the Schwinger effect by cosmological magnetic fields in the early universe~\cite{Kobayashi2021,Kobayashi2022}. The first experimental search for MMs produced by the Schwinger mechanism was carried out by MoEDAL in the ultraperipheral Pb-Pb collisions at the LHC, establishing mass limits up to 75 GeV/c$^2$ at 95\% C.L.~for 1-3 $g_{\mathrm{D}}$ MMs~\cite{MoEDAL_nature}{\textcolor{red}, with more recent searches from both MoEDAL~\cite{MoEDAL_beampipe} and ATLAS~\cite{ATLAS:2024nzp}}. While the sensitivity of such searches will increase during the upcoming heavy-ion runs at the LHC, HL-LHC~\cite{hllhc}, and HE-LHC~\cite{helhc}, our projections show that the probed MM masses are unlikely to exceed 200-300 GeV/c$^2$, which does not reach the range suggested by theoretical models. The projections, whose methodology is described in more details in Ref.~\cite{aditya_projections}, are based on full Monte Carlo (MC) simulation of the relevant physics, realistic detector geometry of a MoEDAL-like detector, and expected luminosity targets. The question we ask is how one could get to the motivated mass region in the next few decades, if it's at all possible.

The paper considers three potential frontiers for next-generation MM searches -- collisions of CRs with the atmosphere, heavy-ion collisions at accelerators, and relics from primordial phase transitions. MMs predicted by the theories cited above most commonly carry magnetic charges from 1 to 3 Dirac units, so this investigation focuses on that range.

\section{Collisions of cosmic rays with the atmosphere}
\begin{figure}[htpb]
 \centering
 \includegraphics[width=0.49\textwidth]{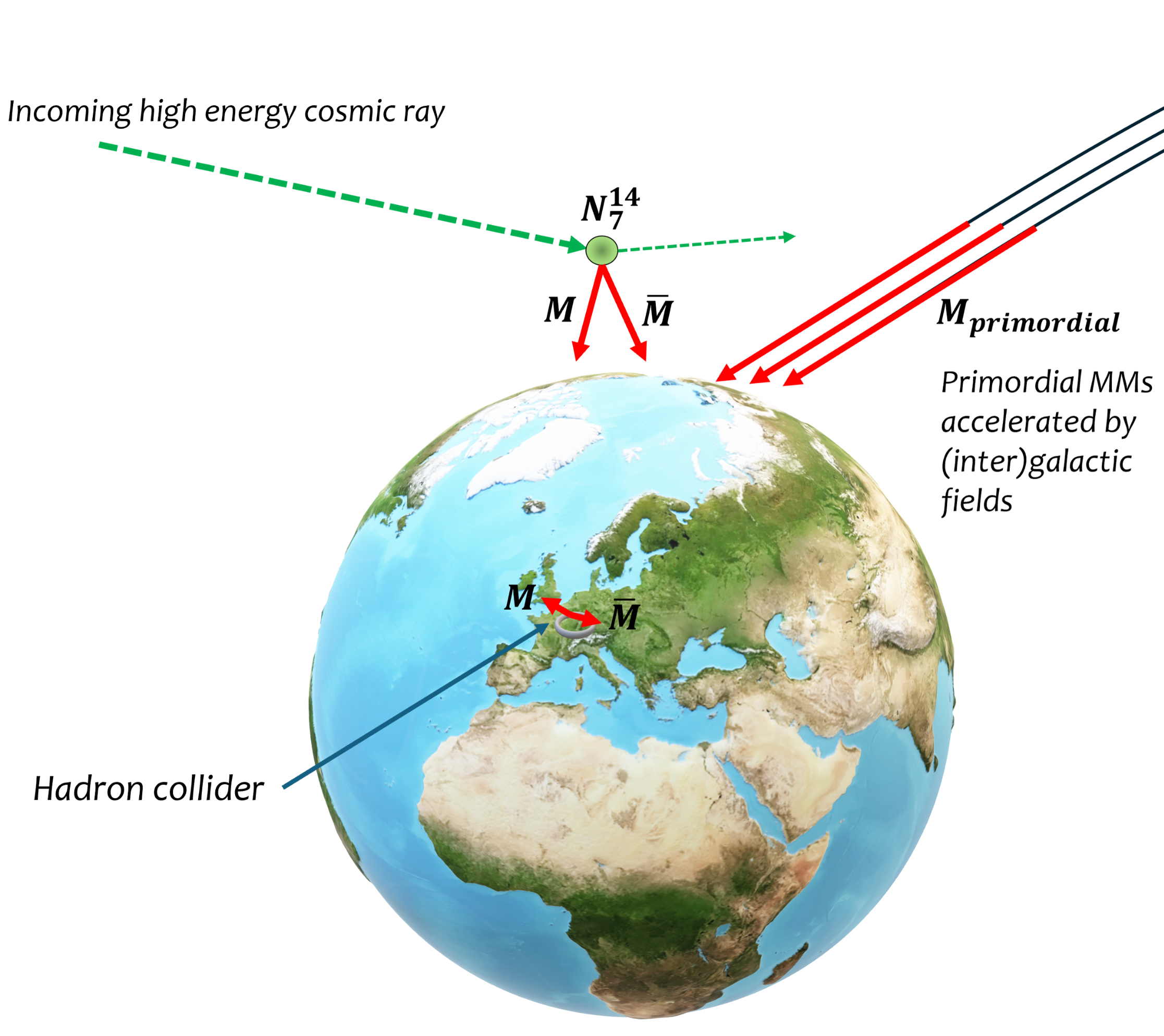}
 \caption{Illustration of the three different MM sources considered in this work: a) MMs produced via the Schwinger process in ultraperipheral collisions of CRs with atmosphere nuclei; b) MMs produced via the Schwinger process in heavy-ion collisions at a hadron collider; c) primordially produced MMs. }
 \label{fig:production}
\end{figure}

Low-mass MMs could be continuously created in the Earth's atmosphere when CRs pass by atmosphere nuclei. Figure~\ref{fig:production} illustrates the concept. 
According to Amp\`{e}re's law, enormous magnetic fields are generated briefly during such flybys, giving rise to the Schwinger production of MMs. The MMs would travel downwards and, depending on their initial momentum, lose all momentum in the atmosphere, or slam into the surface. In the former case, the Earth's magnetic field will start guiding them towards the poles, where they will eventually touch down. In what follows, we describe the approach to calculating the production rate and trajectories of such MMs.  

\begin{figure}[htpb]
\centering
\includegraphics[width=0.49\textwidth]{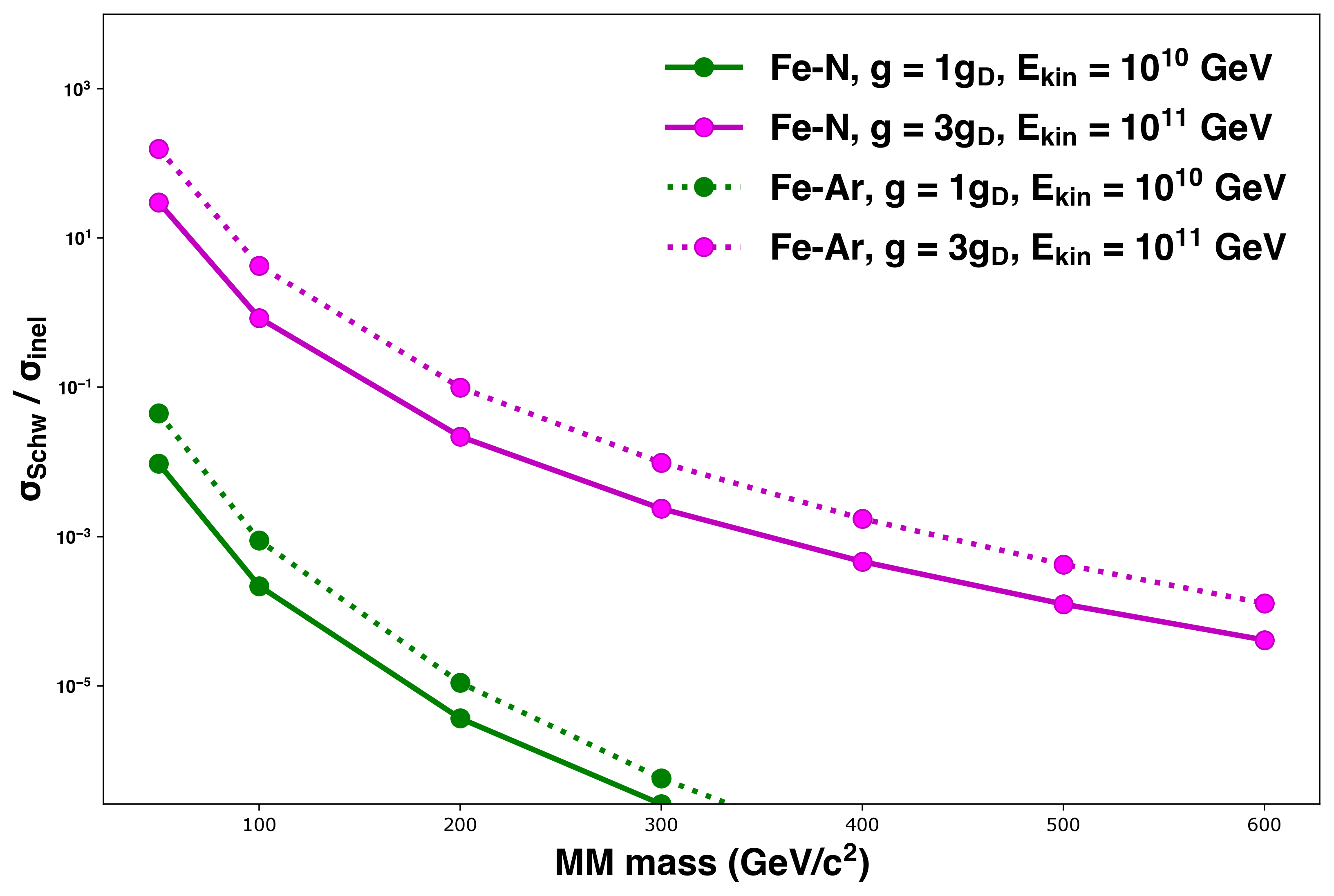}
\caption{The ratio of Schwinger MM production and inelastic scattering cross sections as a function of MM mass. Solid lines correspond to collision of CR iron ions with atmospheric nitrogen, while dash lines correspond to Fe-Ar collisions. Green lines correspond to initial CR energy of 10$^{10}$ GeV and MM with 1 Dirac charge. Magenta lines correspond to CR energy of 10$^{11}$ GeV and MMs with 3 units of Dirac charge.}
\label{fig:inel}
\end{figure}

The production cross section and center-of-mass kinematics are calculated following the formalism developed in Refs.~\cite{arttu_prd2019, Gould:2021bre}. To be conservative, we use the smaller of the two cross section approximations described in the references. The electromagnetic fields, $E$ and $B$, produced by each considered ion are computed by integrating the Li\'{e}nard-Wiechert potentials over classical nuclear charge distributions inferred from elastic scattering~\cite{DeVries:1987atn, Lewin:1995rx, Cui:2021vgm}. Due to the boost from the center-of-mass to the Earth frames, the produced MMs would be highly relativistic (Lorentz factors, or $\gamma$, of up to 10$^5$) and propagate towards the surface. The flux of incoming CRs is calculated using the Global Spline Fit~\cite{Dembinski:2017zsh}, a data-driven model that characterizes the flux and composition from 10 GeV to 10$^{11}$ GeV. The statistical and systematic uncertainties on the flux components described in the reference only marginally affect the MM mass limits, due to the Schwinger production cross section depending exponentially on the mass~\cite{MoEDAL_nature}. Protons are the most abundant component of the CRs but produce a magnetic field with small total volume and energy, limiting the production of composite MMs. We quantify this threshold effect by ensuring that the energy density of the magnetic field integrated over its peak
is greater than the energy of the monopole pair. Consequently, we concentrate on heavier components of CRs, which lead to larger magnetic field energies, in particular the iron ions. Similarly, while nitrogen is the most abundant element in the Earth's atmosphere, in some cases the interaction of CRs with heavier atmospheric elements such as oxygen, argon, or xenon results in higher production rates. The Schwinger production cross section calculated for relevant pairs of colliding ions, MM magnetic charge, and incoming CR energy is then compared to the Standard Model (SM) inelastic cross section (Figure~\ref{fig:inel}). 
The latter accounts for competing processes that could destroy the CR before it produces a MM. We employ a toy MC to evaluate the fraction of such cases. The mean free paths for both interactions used in the random draws depend on the elevation. The atmosphere is approximated by a series of one hundred layers with different average densities and composition, from the surface up to the K\'{a}rm\'{a}n line, that are modeled according to the NRLMSISE standard atmospheric model~\cite{NRLMSISE:2020}. We find that for all relevant MM masses and initial energies, no more than 1 in $\sim$10$^5$ CR ions will produce a MM before experiencing an inelastic process. This estimate conservatively ignores MM production from secondaries, which could yield a few times more MMs, given that the highest fragmentation branching ratios for heavy nuclei with $A$ nucleons are to $A-1$ or $A-2$~\cite{Morejon:2019pfu}. Knowing the fraction of incoming CR ions that produce MMs, we then estimate the expected MM detection rate for different experimental searches discussed later. We find that this channel is only sensitive to MMs with masses $\lesssim 80$~GeV/c$^2$, which are already excluded~\cite{MoEDAL_nature, MoEDAL_beampipe}. It is worth emphasizing that this conclusion contradicts the claim that CR-atmosphere collisions set leading limits on 1-100 TeV/c$^2$ MM~\cite{ryan_2022} because the latter work is not applicable to composite MMs, which are the type predicted by most modern models.

\section{particle collisions at accelerators}
\begin{figure}[htpb]
\centering
\includegraphics[width=0.49\textwidth]{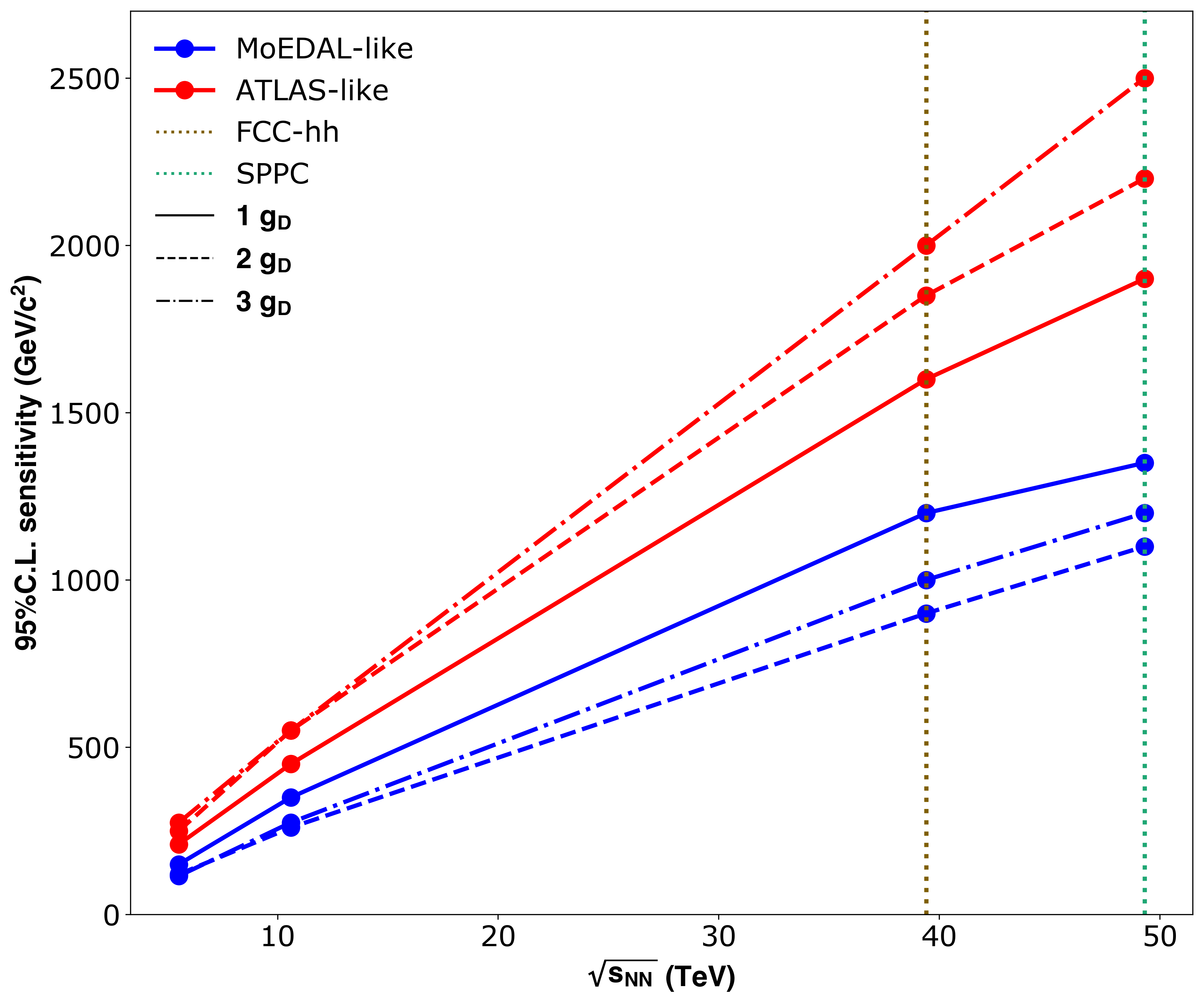}
\caption{Expected sensitivity of MM searches in Pb-Pb collisions as the function of the collision energy. The expected energies for the FCC-hh and SPPC machines are marked by the brown and jade dashed vertical lines, respectively. The blue lines correspond to a MoEDAL-like detector (location farther away from the interaction point, low efficiency), while the red dot-dashed line corresponds to a general-purpose like detector (location close to the interaction point, high efficiency) that is optimistically assumed to have zero background and able to detect multiply charged MMs.}
\label{fig:next_collider}
\end{figure}

The next considered frontier is the hadron collisions at accelerators. Currently, two similar proposals exist for the next-generation hadron collider, the FCC-hh~\cite{fcc} and SPPC~\cite{sppc}. The former is foreseen as a 100 TeV machine tentatively expected to start 40-45 years from now~\cite{fcc_hh_2065}. The latter is expected to reach 125 TeV and begin construction in at least 20 years~\cite{sppc_2044}. Figure~\ref{fig:next_collider} shows projected sensitivity to MMs produced in Pb-Pb collisions via the Schwinger effect at the next collider. 
The calculation assumes the ultimate scenario~\cite{fcc} for the integrated luminosity of Pb-Pb collisions (110 nb$^{-1}$) for both the FCC-hh and SPPC machines and follows the same methodology as in Ref.~\cite{aditya_projections}. Additionally, we anticipate progress in the detector technology and assume that future general-purpose collider detectors will be able to overcome the difficulties with reliable detection and reconstruction of highly ionizing particles~\cite{tpc_high_dedx}, allowing them to combine their high efficiency (assumed 50\% here~\cite{ATLAS2023_MM}) with sensitivity to magnetic charges higher than 1 $g_{\mathrm{D}}$. As the Figure shows, a 100-125 TeV machine will reach sensitivity to TeV/c$^2$ MM masses, addressing some of the existing models.

\section{Primordial monopoles}
\label{sec:primordial}
With the next-generation collider several decades away, we turn our attention to the third frontier -- search for primordial MMs arriving to Earth as high-energy CRs. MMs predicted in theories based on broken symmetries are expected to have been produced in the early universe when the temperature was on the order of the relevant phase transition. The freeze-in of long-wavelength fluctuations through the transition is predicted to produce a finite density of MMs~\cite{Kibble:1976,Zurek:1985,Hindmarsh:2000kd,arttu:2003}, which then do not annihilate efficiently~\cite{Zeldovich:1978}. For the case of heavy, GUT-scale MMs this created the so-called MM problem. However, this conclusion depends on relatively unconstrained early universe cosmology, including inflation and reheating. Crucially, if the reheating temperature is lower than the phase transition temperature, then MMs would not be formed this way. The reheating temperature could be as low as a few MeV, yielding weak constraints on MMs~\cite{Gould:2017zwi}. More recently, it has been realized that another potential source of cosmic MMs is the Schwinger process in primordial magnetic fields~\cite{Kobayashi2021,Kobayashi2022}. Both low-mass and heavy MMs could have been produced by the Schwinger process in the early universe, but this mode of production also suffers from uncertainties, as the properties of the primordial fields are currently not well understood. Consequently, the disadvantage of all searches for primordial MMs is the inability to conclusively exclude the existence of MMs with a given mass, charge in case of a null result. Nevertheless, the possibility of a discovery motivates these searches, especially if existing infrastructure could be exploited to minimize costs. In what follows, we propose three experimental directions and quantify their reach. Using detailed simulations and calculations, we demonstrate that the proposed searches are feasible and will lead to world-leading sensitivities to low-mass MMs during the next few decades. We choose the two staple detector types that are optimized for detecting magnetic charge -- Nuclear Track Detectors (NTDs) and Superconducting Quantum Interference Device (SQUID). The former are inexpensive, allow covering of large areas while having practically zero SM backgrounds~\cite{patrici:2015}. This is because only highly ionizing particles can leave tracks in the NTDs, and in the experimental context only an exotic particle such as a MM is likely to penetrate multiple layers of NTDs simultaneously. Moreover, a magnetically charged particle's ionization density increases with velocity -- the opposite behavior of an electrically charged one. So, a stack of NTDs can allow differentiating between electric and magnetic charges by registering an increase or decrease of ionization density in subsequent layers. The latter method is the most direct and reliable way to identify an isolated magnetic pole bound to baryonic matter~\cite{Milton_2002, Milton_2006, Kittel_PhysRevB.15.333, Goebel1983, Gamberg2000-ms, BRACCI1984236, OLAUSSEN1985465}. This is because any magnetic dipole passing through the magnetometer would induce currents that cancel each other out, whereas a MM would induce a persistent current. The current persists for as long as the SQUID is in a superconductive state, with its value directly proportional to the magnetic charge and independent of the MM's mass. Several instrumental and environmental effects could cause spurious readings in a SQUID magnetometer~\cite{moedal_first}. They could easily be averaged out by repeated measurements, while a MM would consistently yield the same value. SQUIDs have been used in MM searches in the early days~\cite{giacomelli} as well as now~\cite{moedal_first,moedal_prl:2017,moedal_prl2019,moedal_run2_2, HECO_1, HECO_2}. 

\begin{figure}[htpb]
 \centering
 \includegraphics[width=0.49\textwidth]{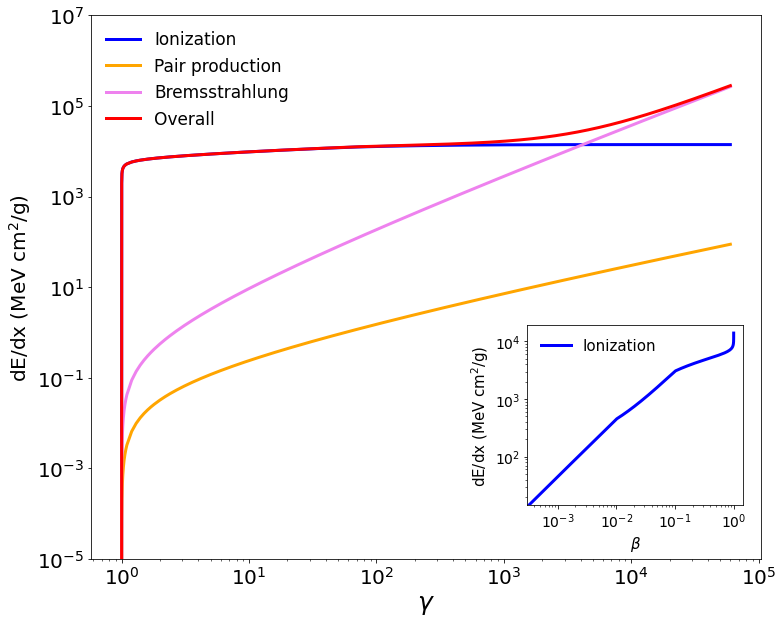}
 \caption{Energy loss of a 100 GeV/c$^2$, 1 $g_{\mathrm{D}}$ MM in the atmosphere as a function of its Lorentz factor $\gamma$, as implemented in Geant4 simulation used in this work. The insert shows the energy loss of slow-moving MMs versus the MM's $\beta$, dominated by ionization (blue). Pair production (yellow) and bremsstrahlung (pink) begin to dominate the total energy loss (red) for Lorentz factors above $\sim$10$^3$.}
 \label{fig:EnergyLoss}
\end{figure}

\begin{figure}[htpb]
 \centering
 \includegraphics[width=0.49\textwidth]{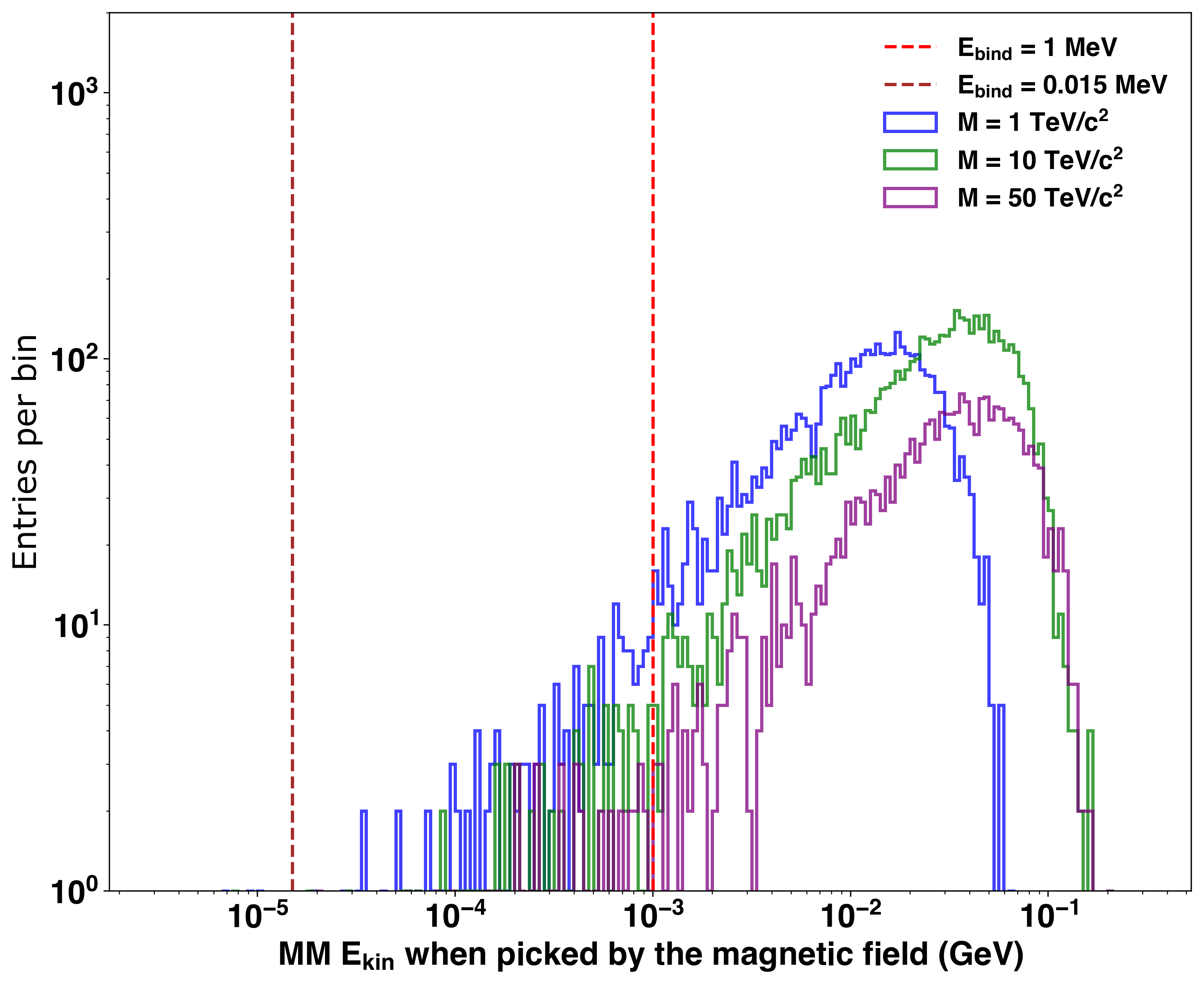}
 \caption{Distribution of kinetic energies of MMs picked up by the Earth's magnetic field for three choices of MM mass: 1 (blue), 10 (green), and 50 (purple) TeV/c$^2$. Dashed vertical lines encapsulate the range of typical binding energies between MMs and atoms.}
 \label{fig:slowedMMs}
\end{figure}

We assume that the flux of MMs produced in the early universe would be isotropic. Given the estimated strength and coherence lengths of (inter-)galactic magnetic fields, MMs lighter than $\lesssim$10$^{7}$ TeV/c$^2$ would not be gravitationally bound to the galaxy and acquire relativistic velocities. A recent investigation allows for a wide range of $\gamma$s of 1 $g_\mathrm{D}$ MMs passing by the Earth~\cite{perri_intergalactic_acc_2023}. To calculate the sensitivity of a given experiment to primordially produced MMs, we first simulate incoming MMs with a given mass, charge, and Lorentz factor when entering the Earth's atmosphere ($\gamma_\mathrm{in}$). The MM physics, transportation, and energy losses are implemented using the Geant4 toolkit~\cite{geant4}. The MM's ionization energy losses are modeled using the formalism described in Refs.~\cite{Ahlen:1978, bologna_2016,Ahlen:1982mx} that provide an accurate description of total energy loss from non-relativistic (down to $\beta$ of 10$^{-3}$) to highly relativistic ($\gamma$ up to 10$^{2}$) MMs. Pair production and bremsstrahlung are implemented as described in Ref.~\cite{WICK2003663} and begin to dominate energy losses at higher energies (see Figure~\ref{fig:EnergyLoss}), with bremmsstrahlung being the largest contributor for the MM masses considered here. Not included is the contribution from the photonuclear effect that competes with pair production at Lorentz factors above 10$^4$ but only weakly affects the results.
The simulation geometry includes the Earth's surface, atmosphere, and an approximate description of the considered experiment. The atmosphere is simulated as described earlier. The Geant4 implementation of the Earth's magnetic field, crucial to simulate the trajectories of low-mass MMs, is based on the \textsc{MAGNETOCOSMICS} model~\cite{Desorgher20056802}. The model includes both the International Geomagnetic Reference Field (IGRF)~\cite{IGRF2021-dr} and external magnetospheric field~\cite{TSYGANENKO}. The latter describes the field's asymmetry due to the solar wind. The force due to the Earth's electric field~\cite{earth_e_field} on MMs can curve their trajectories, but is maximally only a few percent of that due to the magnetic field, so we neglect it. A slowed-down MM may bind to an atmospheric atom, or its constituents, while traveling, which would act as an extra drag. However, the binding energies~\cite{Milton_2006} are typically in range of keV to, at most, an MeV (with aluminum nuclei, not present in the atmosphere above trace levels, being an exception and binding at $\sim$0.5-2.5 MeV). Only a small fraction of MMs slow that much down, as shown on Figure~\ref{fig:slowedMMs}.

\subsection{Primordial monopoles trapped near the Earth's magnetic poles}
\begin{figure}[htpb]
\centering
\includegraphics[width=0.49\textwidth]{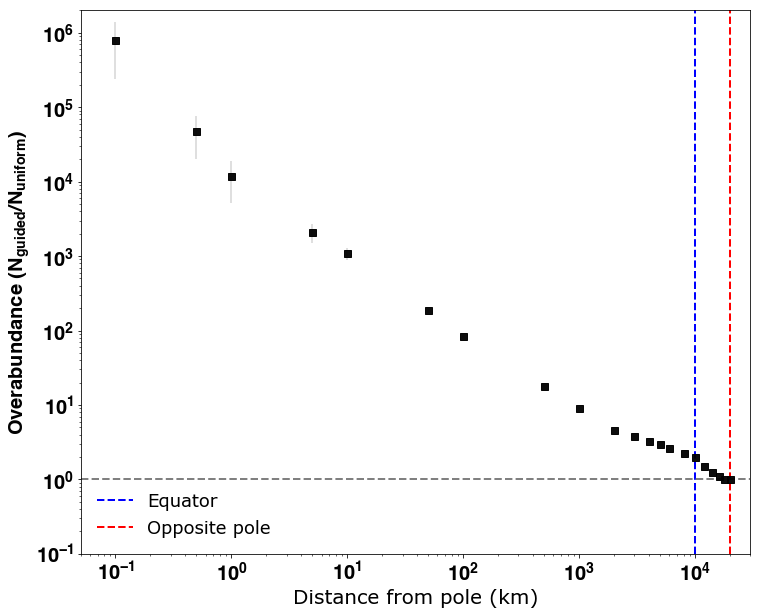}
\caption{The overabundance of MMs touching down near the Earth's magnetic pole vs. the distance from the pole. MMs were simulated descending with initial positions uniformly distributed over the Earth's surface at 100 km altitude. The initial velocity was small enough for MMs to be picked up the Earth's magnetic field. The overabundance for a given distance from the pole is calculated as the number of MMs touching down within the given area, divided by the expected number of touchdowns in the same area assuming a uniform distribution. The blue and red vertical lines correspond to the distance from the pole to the equator and the opposite pole, respectively. The dashed horizontal line represents an overabundance of 1.}
\label{fig:pole_conc}
\end{figure}

The first experiment would take advantage of infrastructure developed at the south pole. As our simulations show, MMs with low $\gamma_\mathrm{in}$ would slow down in the atmosphere and be guided by the Earth's magnetic field towards the magnetic poles (also known as ``dip'' poles), producing an overabundance of touchdowns in these two areas, making them a natural locus for future searches. 
Figure~\ref{fig:pole_conc} shows the increase in the rate of touchdowns near the magnetic poles predicted by the simulation for the low-mass MMs with $\gamma_\mathrm{in}$ of 1 to 1000, depending on the mass. The overabundance of touchdowns, defined as the number of touchdowns within a given distance from the pole in the simulation, divided by the number expected for the same area assuming a uniform distribution. The overabundance shown on the figure decreases steeply away from the dip pole but remains large a substantial distance away from the pole, which is likely because a fraction of the Earth's field lines do not reach the poles and touch the surface earlier. The overabundance corresponding to the distance from the pole to equator (i.e., for the half of the Earth surface) is found to be close to 2, meaning that the vast majority of MMs with initial positions in the opposite hemisphere are guided past the equator.

\begin{figure}[htpb]
 \centering
 \includegraphics[width=0.49\textwidth]{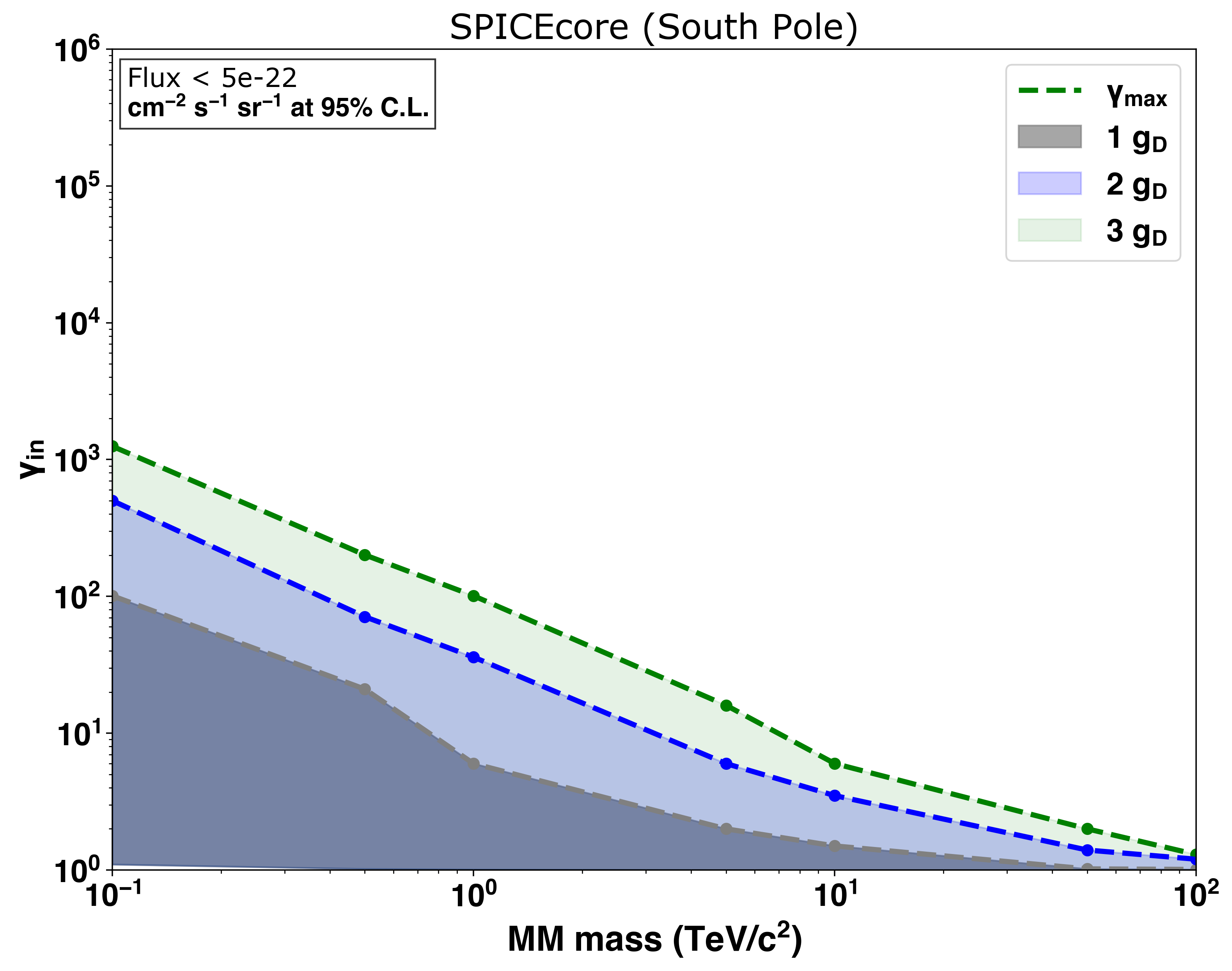}
 \caption{Expected 95\% C.L.~exclusion limits on the flux of cosmic MMs if none are found to be trapped in the SPICEcore samples. Fluxes of MMs with masses and initial Lorentz factors in the shaded regions are excluded. Grey, indigo, and green lines correspond to MMs with 1, 2, and 3 units of Dirac charge, respectively. The dashed lines show the boundaries of the max initial Lorentz factors.}
 \label{fig:ice}
\end{figure}

While the exact location of the Earth's magnetic poles is subject to the geomagnetic secular variation, it coincides with the Earth's geographic poles when averaged over a few thousand years~\cite{average_pole_loc}. Upon hitting the surface, the MMs will lose the remaining kinetic energy within a $\sim$m of ice and get trapped by protons with a binding energy of 15-1000 keV~\cite{Milton_2006}. The ice could then be analyzed by a SQUID magnetometer for an unambiguous signature of an isolated magnetic charge -- the persistent current. The South Pole Ice Core (SPICEcore) project drilled a 1751 m deep core in the ice near the geographic pole~\cite{SPICEcore_2021}. At its maximum depth, the SPICEcore samples date back $\sim$54k years~\cite{SPICEcore_2019}. During that period, the Earth's magnetic field varied considerably~\cite{Constable2007}. In particular, the polarity of the field reversed briefly (for a few hundred years) during the Laschamp excursion roughly 42k years ago. The polarity of the Earth's field does not affect the accumulation of MMs, since they are always created in pairs of opposite polarity. It is estimated that the average virtual axial dipole moment was 10-20\% higher 50k years ago than now, before decreasing by a factor of two 40k years ago, then recovering to the peak value two thousands years ago, and finally decreasing again by 10-20\% to the present value. The magnitude of the Earth's field affects the fraction of MMs that are picked up by the field. However, due to a large disparity between the initial kinetic energy of a MM and the potential energy of the Earth's magnetic field, a factor of two difference in the strength of the latter translates to just several tens of meters of the deceleration path in the atmosphere, so does not affect the results appreciably. The SPICEcore samples are 98 mm in diameter and up to 2 m in length, thus are small enough to pass through SQUID magnetometers used for MM searches~\cite{second_event_magnetometer}. The National Science Foundation Ice Core Facility (NSF-ICF) currently stores 13.2 cubic meters of ice from the drilling activity~\cite{IceCore_Curator}. Scanning the existing samples could be accomplished in about a year. The exact flux limit depends on the value of the overabundance of touchdowns at the drilling location averaged over the age of the samples, which is difficult to calculate accurately. Here, we assume an overabundance that is two orders of magnitude lower than the maximum overabundance corresponding to the exact location of a dip pole, giving the 95\% C.L.~flux limit of approx. $<$5$\cdot$10$^{-22}$ cm$^{-2}$s$^{-1}$sr$^{-1}$ for 1-3 $g_{\mathrm{D}}$ MMs with masses from 0.1 to up to a 100 TeV/c$^2$. This flux limit, shown in Figure~\ref{fig:ice}, is many orders of magnitude below the Galactic Parker bound, $\sim$10$^{-16}$ cm$^{-2}$s$^{-1}$sr$^{-1}$~\cite{perri_intergalactic_acc_2023}. It is also substantially below the seed Galactic Parker bound ($\sim$10$^{-17}$ -- 10$^{-21}$ cm$^{-2}$s$^{-1}$sr$^{-1}$ for different assumed values of the intergalactic magnetic field strength~\cite{perri_intergalactic_acc_2023}), whose value has recently been updated, showing that its value could be relaxed by orders of magnitudes for fast-moving MMs, compared to the earlier estimate made for slow, i.e., heavy MMs. 
It is stronger than limits from other experiments~\cite{pdg_2022,Auger2016} but only applies to MMs with $\gamma_\mathrm{in}$ not exceeding the value of 1 to 1000, depending on the MM's mass. Specifically, the following results, summarized in Refs.~\cite{pdg_2022, Auger2016}, that currently dominate the field will be surpassed: the Pierre Auger Observatory has set flux limits on MMs with masses as low as 0.1 TeV/c$^2$ from $<$8.4$\cdot$10$^{-18}$ to $<$2.5$\cdot$10$^{-21}$ for $\gamma_\mathrm{in}$ exceeding 10$^8$ to 10$^{12}$, respectively~\cite{Auger2016}; RICE~\cite{RICE_PRD} and ANITA~\cite{ANITA_PRD} reached limits on the order of 10$^{-19}$ -- 10$^{-20}$ but were only sensitive to $\gamma_\mathrm{in}$ exceeding 10$^{7-8}$ and 10$^{9}$, respectively; ANTARES~\cite{antares_2022} placed upper limits on the flux ranging from $<$1$\cdot$10$^{-17}$ to $<$7.3$\cdot$10$^{-19}$ for $\gamma_\mathrm{in}$ from $\sim$1.2 to $\sim$10, respectively, but only for MM with masses exceeding 10$^8$ TeV/c$^2$, so not relevant for this work; IceCube reported a limit~\cite{icecube_2013} of 3.4$\cdot$10$^{-18}$ that applies to MMs with masses above 10$^{3}$ TeV/c$^2$, so also not relevant for this work; its more recent results is claimed to be sensitive to all masses and sets the limit of $<$2$\cdot$10$^{-18}$, but only applies to a narrow range of $\gamma_\mathrm{in}$ from $\sim$1.6 to $\sim$10~\cite{icecube_2022}. All of the above limits are reported at 90\% C.L. in units of cm$^{-2}$s$^{-1}$sr$^{-1}$ for MMs with 1 unit of Dirac charge. 

The proposed experiment could be organized quickly, only requires a modest investment, and uses a detection technique that produces an unambiguous, background-free signature of magnetic charge. Other polar ice projects could also be included if possible. Notably, the Vostok ice core project has accumulated $\sim$40 m$^3$ of samples~\cite{vostok_ice_core}. While less sensitive due to being extracted farther away from the geographic pole, these samples are interesting due to dating as far back as 420k years, so averaging over several cycles of the Earth's magnetic field variation. Finally, an experiment could extract small ice core samples from the exact historical locations of the geomagnetic south pole.
The locations are known by direct measurements since 1909 and remain on-shore until early 1960s. The samples need to be extracted from only a couple of meters depth. While each sample would correspond to roughly one year exposure, the overabundance of touchdowns for MMs for small $\gamma_\mathrm{in}$ would be equivalent to more than 10$^4$ yrs of accumulation at a location far away from the dip poles.

\subsection{Primordial monopoles as high-energy cosmic rays}
Another approach is deploying a much larger array of NTDs than was used by previous experiments, such as SLIM (427 m$^2$~\cite{SLIM_2009}) and MACRO ($\sim$1200 m$^2$~\cite{MACRO_final}). Similar proposals were made earlier~\cite{cosmic_moedal_100k, cosmic_moedal_2022}, aiming to place the array on a mountain substantially above the sea level to improve sensitivity to low masses and $\gamma_\mathrm{in}$. Since such a placement is challenging and expensive, we first consider a ground-level NTD array, with an 50000 m$^{2}$ coverage area, comparable to the cited proposals. 

\subsubsection{Sea-level detector placement}
\begin{figure}[htpb]
 \centering
 \includegraphics[width=0.49\textwidth]{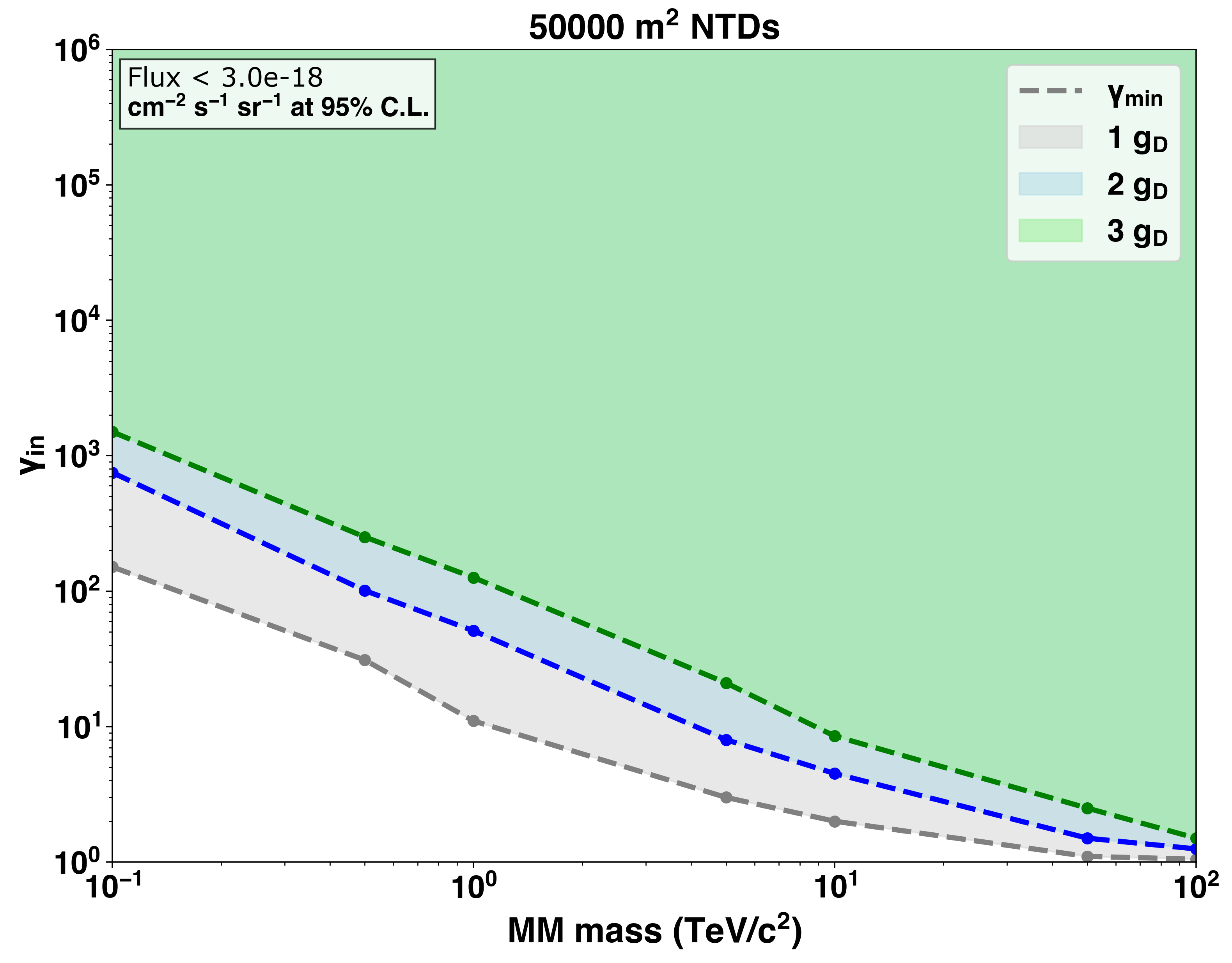}
 \caption{Expected 95\% C.L.~exclusion limits on the flux of cosmic MMs if none are detected by the 50k m$^2$ NTD array after a 10-year exposure. Fluxes of MMs with masses and initial Lorentz factors in the shaded regions are excluded. Grey, indigo, and green lines correspond to MMs with 1, 2, and 3 units of Dirac charge, respectively. The dashed lines show the boundaries of the minimum initial Lorentz factors.}
 \label{fig:ntds}
\end{figure}

Assuming a 10-year exposure, the expected 95\% C.L.~flux limit is $<$3.0$\cdot$10$^{-18}$ cm$^{-2}$s$^{-1}$sr$^{-1}$ for 1-3 $g_{\mathrm{D}}$ MMs with masses from 0.1 to up to a 100 TeV/c$^2$. Complementary to the SPICEcore proposal, the limit for this frugal and more realistic option applies to a region of $\gamma_\mathrm{in}$ $exceeding$ values of $\sim$10$^0$ to $\sim$10$^3$, depending on the mass (Figure~\ref{fig:ntds}). It surpasses current results, such as by the Pierre Auger Observatory~\cite{Auger2016}, IceCube~\cite{icecube_2013,icecube_2022}, ANTARES~\cite{antares_2022}, RICE~\cite{RICE_PRD}, ANITA~\cite{ANITA_PRD}, and others~\cite{pdg_2022} in the $\gamma_\mathrm{in}$ range spanning more than six decades, i.e., from $\sim$12 to 10$^{7-9}$, depending on the mass.

\subsubsection{Polar region detector placement}
An NTD array can also cover the full region of Lorentz factors and masses if placed as close to the present location of the Earth's magnetic pole as practical. It would additionally collect the contribution from MMs that slowed down in the atmosphere and were transported by the Earth's magnetic field. This addition, like with the SPICEcore, would improve sensitivity to MMs with low masses and $\gamma_\mathrm{in}$. The closest permanent stations that could provide support (in order of increasing distance from the current location of the south magnetic pole) are the French's Dumont d'Urville Station, Russian's Vostok Station, and the U.S. South Pole Station, the latter two located at the geomagnetic and geographic south poles, respectively. Both the lower-energy MMs guided by the Earth's magnetic field and higher-energy ones that impact directly above the deployment would pass through several layers of NTDs, producing a characteristic signature. The antarctic placement still suffers from being remote and associated high costs. At the same time, the region already hosts large experiments requiring a much more complicated detectors and infrastructure, such as IceCube~\cite{icecube_design}, BICEP array~\cite{bicep}, and others. The flux limit for the region of small $\gamma_\mathrm{in}$ is estimated to be $<$2.0$\cdot$10$^{-23}$ cm$^{-2}$s$^{-1}$sr$^{-1}$ for 1-3 $g_{\mathrm{D}}$ MMs with masses from 0.1 to up to a 100 TeV/c$^2$. In case of a candidate event, the material below the specific NTD stack could be investigated for the presence of a stopped, bound MM using a magnetometer. 

While NTDs are inexpensive and require essentially no maintenance during deployment, a big challenge with all proposed large-area NTD deployments is the time and cost needed to chemically etch, scan, and analyze such large areas. A promising way to alleviate this is by adding layer(s) of dedicated electronic detectors that are, like the NTDs, inexpensive and sensitive only to highly ionizing particles~\cite{ssbc}. Segmented in a way similar to the NTD sheets, they could pinpoint the location of candidate events, drastically reducing the NTD area that needs to be processed. Even in the case of a polar or mountain deployment, such a detector would not complicate maintenance and operation significantly, as it requires no high voltage, amplifier, or other complicated electronics. The analysis step could be further sped up by emerging machine learning techniques~\cite{Bevan:2019upi}.

\subsubsection{Placement at a cosmic ray observatory}
An important potential placement for an array of NTDs is at a cosmic ray observatory, such as the Pierre Auger Observatory, Large High Altitude Air Shower Observatory, or Telescope Array Project. Some of the detected ultra-high energy cosmic rays (UHECRs), defined as CRs with an energy greater than 1 EeV, do not have trajectories pointing back to any plausible astrophysical sources~\cite{Amaterasu_2023} and have energies larger than what could be explained by the known acceleration mechanisms and what is possible for known particles of remote, intergalactic origin~\cite{gzk_1,gzk_2,heavy_nucl_disint_1969}. It has long been suggested that UHECRs are primordial low-mass MMs~\cite{crs_are_mms_1996} because they are expected to be accelerated to similarly large energies by the intergalactic and galactic magnetic fields and have trajectories not pointing back to specific sources. The recent detection of the Amaterasu particle~\cite{Amaterasu_2023} has reinvigorated such discussions~\cite{Cho_2024,Frampton:2024}. While the Pierre Auger Observatory has published a MM search~\cite{Auger2016}, the experiment is not directly sensitive to magnetic charge and relies on understanding of the MM's air shower profile, which is subject to model-dependent uncertainties. In contrast, placing large arrays of NTDs or other detectors that are reliably sensitive to magnetic charge on the territory of a cosmic ray observatory could directly check the hypothesis of the MM origin of UHECRs. An UHECR's shower core can be located by the observatory with a 50 m resolution~\cite{auger_core_location}, corresponding to the ground area of $\sim$8000 m$^2$. A 50000 m$^2$ NTD array would then be sufficient to provide a coincidence measurement, with the UHECRs' reconstructed position serving as a definitive trigger for the NTD scan. Such an array would only cover a small fraction of the total surface area monitored by the observatory. However, given the measured flux of UHECRs~\cite{Auger_Science_2017}, the expected rate of UHECRs with $>$4 EeV detected in coincidence with the NTD array would be 1 every 5 to 6 years. 

\subsection{Primordial monopoles trapped in mined ore}
\begin{figure}[htpb]
\centering
\includegraphics[width=0.49\textwidth]{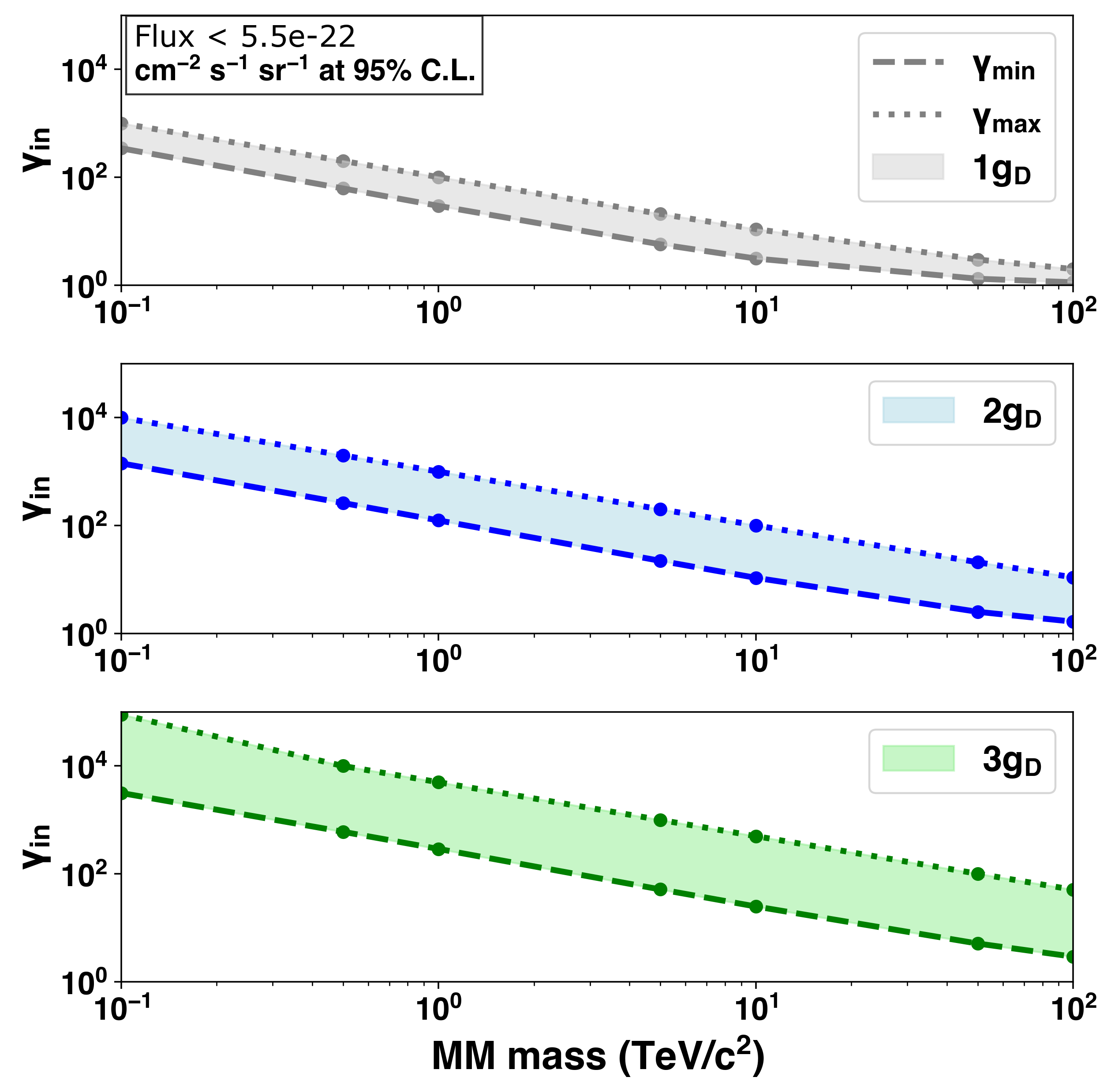}
\caption{Exclusion limits on cosmic MM flux as a function of mass and initial Lorentz factor. The limits correspond to a null result of scanning 1k tons of iron ore deposits in Vale, Brazil. The shaded regions are excluded. The dashed (dotted) lines show the boundaries of the min (max) initial Lorentz factors. Grey, blue, and green colors corresponds to MMs with 1, 2, and 3 units of Dirac charge. Deposits of iron and aluminum ore in other locations could allow similar limits for different ranges of initial Lorentz factors, depending on the depth of the deposits.}
\label{fig:ore}
\end{figure}

Lastly, we consider bringing the earlier searches for MMs trapped in the Earth-based rocks~\cite{PolarRock_PRL, Goto:1963zz, Rocks_PRL_1995}, deep sea sediments, meteorites and lunar rock~\cite{DeepSea71_PRD, DeepSea69_PRL, LunarPRD_1971, LunarPRD_1973} to the next level by exploiting industrial capabilities. Slowed-down MMs are expected to bind to iron and aluminum nuclei with large binding energies~\cite{Kittel_PhysRevB.15.333,Milton_2006}. The production of these metals is currently performed on a vast scale, with $\sim$1M tons of raw ore processed by some facilities every year. In a typical factory, crushed ore is transported by one or more conveyors at speeds of up to several meters per second for up to 5k annual operating hours each year~\cite{conveyor}. The operating company may allow installing a bypass equipped with one or more SQUID magnetometers through which just a small fraction of the total ore would pass, perhaps motivated by publicity and outreach considerations. To estimate the sensitivity of this approach, we consider one concrete example. 
The iron ore deposits at the Carajas-Serra Norte mine in Vale, Brazil, are  ca. 1590 Mtons~\cite{Vale_TechReport} of grade higher than 64\% Fe and estimated exposure time of 2.7G years~\cite{TRENDALL1998_Carajas}. Based on our calculations, processing of 1k tons of iron ore per year (just $\sim$0.001\% of total processed by the company) for one year will result in 95\% C.L.~flux limits of $<$5.5$\cdot$10$^{-22}$ cm$^{-2}$s$^{-1}$sr$^{-1}$ for 1-3 $g_\mathrm{D}$ MMs. The limits, shown on Figure~\ref{fig:ore}, are extremely strong but apply to the specific range of $\gamma_\mathrm{in}$, dictated by the location of the deposits and MMs' energy losses. Other mines, e.g., the Weipa bauxite mine, Mount Whaleback, and Sischen iron ore mines, would provide similar sensitivity for different ranges of $\gamma_\mathrm{in}$ depending on the depth of the deposits.

While not all of the previous similar studies cited above report flux limits, they are based on, at most, a few hundred kilograms of material with exposure times ranging from a few million years for Earth-based rocks to at most 3.6G years for the extraterrestrial ones. We comment on the older studies cited in this section and provide our own estimates of the flux limits that they reached in the Appendix. Overall, in the absence of additional accumulation effects that apply to these studies, like the one near dip poles discussed earlier, the proposed experiment will by design offer a drastic increase in the sensitivity. The reason why some of the previous studies do not report flux limits is likely due to difficulty of calculating it reliably, given the lack of pertinent history of the samples. In contrast, the geological properties of the ore deposits discussed here are well-studied, allowing one to reliably identify the range of masses and Lorentz factors of MMs potentially trapped there. Additionally, some of the previous-generation studies in this frontier relied on the assumption that trapped MMs could be extracted by an external magnetic field with the strength on the order of a few Tesla~\cite{DeepSea71_PRD, DeepSea69_PRL}, which has since been refuted~\cite{Milton_2006}. 

\section{Concluding remarks}
To summarize, the proposed experiments would provide world-leading sensitivities to low-mass MMs until the next hadron collider turns on, two to four decades from now. While null results would not be as informative as that from a Schwinger production experiment at a collider, they offer the best chance of a discovery at a small fraction of the cost and effort, bringing the long-standing quest for isolated magnetic charge closer to completion. 

This study has focused on the unique electromagnetic interactions of MMs. For composite MMs, there are model-dependent non-electromagnetic interactions which are typically limited to the region of the MM core, and which may yield additional signatures.
The cross section calculation used in this work is also known to be conservative. Future development in this area is likely to strengthen the projections. 

A key point is that this work goes beyond setting flux limits based on indirect observations by non-dedicated experiments. The MM production cross section for the Schwinger process is calculable nonperturbatively and is not subject to the exponential suppression for composite MMs, while the proposed primordial MM detection efforts are unambiguously sensitive to magnetic charge and are essentially background-free. Consequently, even a negative result of the proposed collider searches would reliably exclude the existence of MMs with specific masses and charges, while the results of cosmic searches, especially conducted in conjunction with the CR observatories, could confirm or refute the suggestion that UHECRs are MMs. 

\paragraph{Acknowledgments.} This work is supported by the NSF grant 2309505 and by a Dorothy Hodgkin Fellowship from the Royal Society. We thank Ryan Plestid and Marcos Santander for valuable discussions during the early stages of this work. Ostrovskiy thanks Chen Zhang for answering questions about monopoles in cosmology. Upreti thanks Curtis La Bombard for details and clarifications about the NSF-Ice Cores.

\paragraph{Author Contributions.} Simulations, statistical analysis, results, and figures were produced by A. Upreti. Theoretical calculations of the Schwinger production rates were done by O. Gould who also helped edit the manuscript. I. Ostrovskiy conceived and supervised the project, wrote and edited the manuscript. All authors have read and agreed to the final version of the manuscript.

\paragraph{Data Availability Statement.} Data supporting this study is available upon request. The code for computing the electromagnetic fields of ion pairs is available at~\cite{emions}. The code for generating the sensitivity plots is available at~\cite{projBox}. 

\appendix
\section{Comments on the the earlier searches for MMs trapped in the Earth-based rocks, deep sea sediments, meteorites and lunar rock.}

Here we comment on the earlier searches for MMs trapped in the Earth-based rocks~\cite{PolarRock_PRL, Goto:1963zz, Rocks_PRL_1995}, deep sea sediments, meteorites and lunar rock~\cite{DeepSea71_PRD, DeepSea69_PRL, LunarPRD_1971, LunarPRD_1973} mentioned in the paper and interpret their findings in terms of the flux limits. It should be noted that, in the absence of pertinent information, it is not clear to what range of Lorentz factors each limit corresponds. In all likelihood, the relevant range is different for different studies. Given this, we do not think that there is a sound way to combine such estimates for the different searches.

\begin{itemize}

\item Ref.~\cite{PolarRock_PRL}

The mantle rocks could not have accumulated low-mass MMs on the geological time-scale ($\sim$Gyr) because the mantle is too deep for such MMs to reach. The overall mass of the polar rocks considered in the study was 23.4 kg. As described in the reference, the mantle rock samples have been collected from various locations on Earth with prominent volcanic activity, such as Hawaii, Iceland, etc, and thus do not benefit from the MM accumulation factor over the poles. The samples were primarily made of basaltic rock with an average density of 3 ton/m$^{3}$, so the resulting exposure volume is $\sim$7e-3 m$^{3}$, roughly equivalent to a surface exposure area of 4$\cdot$10$^{-2}$ m$^{2}$. The time frame when the basaltic rock was pushed out to the Earth's surface is not known precisely. However, an estimate for these samples puts it at 1-5 million years, during which they would be exposed to low-mass monopoles. While the reference does not provide an estimate of the excluded flux, our naive estimate results in general flux limits of $\sim$ 4$\cdot$10$^{-17}$ (1 Myr) to 8$\cdot$10$^{-18}$ (5 Myr) cm$^{-2}$s$^{-1}$sr$^{-1}$. It should be noted that it is not clear to which range of Lorentz factors and MM masses this would apply. 

\item Refs.~\cite{Goto:1963zz,DeepSea69_PRL,DeepSea71_PRD} are excluded, because it was shown that a 5T field would not extract MMs~\cite{Milton_2006}.

\item Ref.~\cite{Rocks_PRA_1986} used rocks buried at a depth of 25 km, so could not have accumulated low-mass MMs.

\item Ref.~\cite{Rocks_PRL_1995} used 331 kg material (including 112 kg of meteorites) with expected exposure of $~$1G years. This translates to a naive estimate of 6$\cdot$10$^{-21}$ cm$^{-2}$s$^{-1}$sr$^{-1}$ flux limit for an unspecified range of Lorentz factors. It's weaker than the ore limit due to much lower mass and than the SPICEcore limit due to the combination of the lower mass and absence of the pole accumulation effect.

\item Ref.~\cite{LunarPRD_1971} uses 8 kg of lunar samples and assumes the age of crystallization of 3.6G years. This translates into a naive flux limit 2$\cdot$10$^{-20}$  cm$^{-2}$s$^{-1}$sr$^{-1}$. This is weaker than the ore limit due to much lower mass and than the SPICEcore limit due to the combination of the lower mass and absence of the pole accumulation effect.

\item Ref.~\cite{LunarPRD_1973} uses Apollo 11, 12 and 14 samples totaling 11.5 kg. The exposure ages of the samples are 0.5G, 0.36G, and 0.43G years. The naive combined limit is 1.1$\cdot$10$^{-19}$ cm$^{-2}$s$^{-1}$sr$^{-1}$.

\end{itemize}

\bibliography{main}

\end{document}